# Robust Formation of Ultrasmall Room-Temperature Neél Skyrmions in Amorphous Ferrimagnets from Atomistic Simulations

**Chung Ting Ma[1], Yunkun Xie[2], Howard Sheng[3], Avik W. Ghosh[1,2], and S. Joseph Poon[1*]**

[1]Department of Physics, University of Virginia, Charlottesville, Virginia 22904 USA
[2]Department of Electrical and Computer Engineering, University of Virginia, Charlottesville, Virginia 22904 USA
[3]Department of Physics and Astronomy, George Mason University, Fairfax, Virginia 22030 USA
[*] sjp9x@virginia.edu

Neél skyrmions originate from interfacial Dzyaloshinskii Moriya interaction (DMI). Recent studies have explored using thin-film ferromagnets and ferrimagnets to host Neél skyrmions for spintronic applications. However, it is unclear if ultrasmall (10 nm or less) skyrmions can ever be stabilized at room temperature for practical use in high density parallel racetrack memories. While thicker films can improve stability, DMI decays rapidly away from the interface. As such, spins far away from the interface would experience near-zero DMI, raising question on whether or not unrealistically large DMI is needed to stabilize skyrmions, and whether skyrmions will also collapse away from the interface. To address these questions, we have employed atomistic stochastic Landau-Lifshitz-Gilbert simulations to investigate skyrmions in amorphous ferrimagnetic GdCo. It is revealed that a significant reduction in DMI below that of Pt is sufficient to stabilize ultrasmall skyrmions even in films as thick as 15 nm. Moreover, skyrmions are found to retain a uniform columnar shape across the film thickness due to the long ferrimagnetic exchange length despite the decaying DMI. Our results show that increasing thickness and reducing DMI in GdCo can further reduce the size of skyrmions at room temperature, which is crucial to improve the density and energy efficiency in skyrmion based devices.

## Introduction

Magnetic skyrmions have topologically protected spin textures. Their potential in advancing memory density and efficiency has drawn extensive investigation in recent years[1-25]. In magnetic materials, skyrmions are stabilized through the Dzyaloshinskii Moriya interaction (DMI)[26-27], generated by either inherent chiral asymmetries or by interfacial symmetry breaking. Intrinsic DMI arises in non-centrosymmetric crystals such as B20 alloys, where Bloch skyrmions have been found in MnSi and FeGe at low temperature[12-13]. On the other hand, interfacial DMI originates from inversion symmetry breaking by an interfacial layer with strong spin-orbit coupling. Multilayer stacks, such as Pt/Co/Os/Pt, Ir/Fe/Co/Pt and Pt/Co/Ta, have been found to host 40 nm to over 1 µm Neél skyrmions at room temperature[14-16]. Several challenges remain in developing skyrmion based memory and logic devices - for instance, skyrmion Hall effect can present a significant challenge in guiding skyrmions linearly along racetracks [19-22]. More critically, aggressive reduction in skyrmion sizes is needed to optimize skyrmion based devices, whereupon their room temperature stability becomes a problem. Thicker magnetic layers are required in most cases to increase stability[17-18]. However, for ferromagnet (FM)/heavy metal multilayer stacks, increase in thickness of the magnetic layer can lead to a loss of interfacial anisotropy and the reduction of the strength of average DMI[28-31], both of which are critical for skyrmion formation. To overcome these challenges, we need to consider a suite of materials and understand their underlying physics, especially with varying film thickness.

Amorphous rare-earth-transitional-metal (RE-TM) ferrimagnets (FiM) are potential candidates to overcome these challenges. Several properties of RE-TM alloys provide a favorable environment to host small skyrmions at room temperature. Their isotropic structure helps with avoiding defect pinning[18], while their intrinsic perpendicular magnetic anisotropy (PMA) [32-35] helps stabilize small skyrmions by allowing the use of thicker films (> 5 nm). However, the effectiveness of interfacial DMI decreases significantly away from the interface[28-31], which is the focus of our present investigation. Besides PMA, the magnetization of RE-TM alloys vanishes at the compensation temperature[36]. With near zero magnetization and near compensated angular momentum[37], the skyrmion Hall effect is vastly reduced[18-20,23], and the skyrmion velocity is predicted to be maximum near the compensation point of angular momentum[61]. Recently, all-optical helicity dependent ultrafast switching has been demonstrated in RE-TM alloys using a circularly polarized laser[38-45]. This gives an additional tool to control spins in device structures. Indeed, RE-TM alloys have begun to draw interest in the field of skyrmion research. Large skyrmions of ~150 nm have been observed in Pt/GdFeCo/MgO[23], and skyrmion bound pairs have been found in Gd/Fe multilayers[24]. Recently, small skyrmions approaching 10 nm were found in Pt/GdCo/$TaO_x$ films[25]. Understanding these results will enable further tuning to reduce the size of the skyrmions. To guide experiments, numerical simulation has served as an important tool, especially for complex systems such as RE-TM alloys[40,46-50]. Several methods, such as atomistic Landau-Lifshitz-Gilbert (LLG) algorithm[40,46-49] and micromagnetic Landau-Lifshitz-Bloch (LLB) algorithm[49] have been employed to provide deeper understanding of the magnetic properties in RE-TM alloys.

In this study, an atomistic LLG algorithm[40,46-48] is employed to investigate the properties of skyrmions in GdCo with interfacial DMI. Although the sign of DMI at FM/heavy metal interface is well studied[51-57], the sign of DMI involving a FiM remains complex and is rarely discussed. Here, we consider two scenarios for the DMI between Gd and Co ($d_{Gd\text{-}Co}$). First, DMI between the antiferromagnetic (AFM) pair is set to the same sign as DMI between ferromagnetic pair, i.e. $d_{Gd\text{-}Co} > 0$. Second, the case of $d_{Gd\text{-}Co} < 0$ is considered. The latter appears to be favored by the sign of AFM interaction[27]. Moreover, to incorporate DMI being an interfacial effect, an exponentially decaying DMI is utilized. Simulation results find that with a switched DMI sign, near 10-nm skyrmions remain robust in GdCo films as thick as 15 nm at room temperature. Through numerical tomography maps, we find that skyrmions at room temperature are distributed as a near uniform column in thicker samples, despite a spatially decaying DMI.

## Results and Discussion

We will now begin to investigate if ultrasmall skyrmions in GdCo can survive an exponential DMI reduction over thick sample sizes. To incorporate the amorphous nature of GdCo, we employ an amorphous structure of $RE_{25}TM_{75}$ from ab initio molecular dynamics calculations, as shown in **Fig. 1.** As shown in **Fig. 2**, at 300 K, the magnetization of amorphous $Gd_{25}Co_{75}$ is 5 x $10^4$ A/m, and it has a compensation temperature near 250 K. We begin with an exponential DMI decay away from the interface, as shown in **Fig. 3**. The DMI value discussed herein is the interfacial DMI $D_0$. The decay length of DMI is based on both previous simulations and experiments. DMI calculation in Co/Pt interface finds a significant decrease in DMI beyond the second layer of Co from the Pt interface[30]. Experimental results also find similar decay in Co-Alloy/Pt interface[52-54]. In amorphous GdCo, we adapted the "second layer" as decay length for direct comparison with these findings.

A range of interfacial DMI values, from $d_{Co\text{-}Co} = 0.1 \times 10^{-22}$ J to $d_{Co\text{-}Co} = 2.0 \times 10^{-22}$ J (D = 0.12 to 2.38 $mJ/m^2$), and three thicknesses, 5 nm, 10 nm, and 15 nm, are considered. Only those show skyrmions are shown herein. To shorten computational time, thicker samples of 10 nm and 15 nm are simulated using a 5 nm thick sample by conserving DMI energy density across the film. To conserve the total DMI energy, a faster decay is employed in a 5 nm sample to simulate 10 nm and 15 nm thick samples to keep the sum of DMI energy to be the same. To check the validity of this simplification, we have compared the results of 10 nm thick samples and 5 nm thick samples with faster DMI decay to verify that the two sets of samples produce identical results. First, we consider two scenarios for the sign of $d_{Gd-Co}$, as both + and - signs have been reported in antiferromagnetically coupled systems[59,60]. **Fig. 4** shows the color maps of equilibrium spin configurations at 300 K for both $d_{Gd-Co} > 0$ and $d_{Gd-Co} < 0$ . For the case of $d_{Gd-Gd}$, $d_{Co-Co} > 0$ and $d_{Gd-Co} > 0$, the simulation with $d_{Co-Co} = 0.25 \times 10^{-22}$ J, $d_{Gd-Gd} = 2.96 \times 10^{-22}$ J and $d_{Gd-Co} = 0.86 \times 10^{-22}$ J corresponds to an average DMI of D = 0.21 mJ/m$^2$. The value of $d_{Gd-Gd}$ and $d_{Gd-Co}$ is calculated from $d_{Co-Co}$ by multiplying the ratio of Gd moment $\mu_{Gd}$ over Co moment $\mu_{Co}$. Further discussion in the supplementary material shows that ***for a given average DMI, the energy minimum and thus skrymion size*** is independent of how each DMI term varies. **Eq. 3** shows the formula used for converting atomistic DMI to average DMI for $Gd_xCo_{1-x}$.

$$\mathrm{D} = \frac{2}{\pi}\frac{1}{\bar{n}}\left[(1-x)\left(\frac{\overline{n_{Co-Co}}d_{Co-Co}}{\overline{r_{Co-Co}}^2} + \frac{\overline{n_{Co-Gd}}|d_{Gd-Co}|}{\overline{r_{Co-Gd}}^2}\right) + \mathrm{x}\left(\frac{\overline{n_{Gd-Gd}}d_{Gd-Gd}}{\overline{r_{Gd-Gd}}^2} + \frac{\overline{n_{Gd-Co}}|d_{Gd-Co}|}{\overline{r_{Gd-Co}}^2}\right)\right] \quad (1)$$

Where $\bar{n}$ is the average number of nearest neighbors around all atoms, $\overline{n_{A-B}}$ is the average number of atoms A that are nearest neighbors to atom B, $\overline{r_{A-B}}$ is average distance between atoms A and nearest neighboring atom B. The $\frac{2}{\pi}$ factor comes from averaging of the cross product $\boldsymbol{s_i} \times \boldsymbol{s_j}$ in DMI energy.

For 5 nm GdCo, with $d_{Gd-Co} < 0$, $d_{Co-Co} < 0.25 \times 10^{-22}$ J, only ferrimagnetic states are observed. At $d_{Co-Co} > 1.0 \times 10^{-22}$ J, skyrmions are elongated due to boundary effect in the simulation or stripes states are observed. The range of DMI, where skyrmions are found, is smaller compared to calculation by Cort *et al.*[21]. This is due to a reduction in anisotropy and exchange stiffness in GdCo. With less DMI energy required to create skyrmions, smaller DMI value is needed to create skyrmions and stripes in FiM. Furthermore, experiment results have measured DMI value greater than 1 mJ/m$^2$ only at ordered FM/heavy metal interface[50-56]. The DMI value at amorphous FiM/heavy metal remains unknown. Due to disorder nature of amorphous materials, the DMI value in amorphous FiM can be much smaller than the DMI value observed in ordered FM.

As shown in **Fig. 4**, with ferromagnetic DMI ($d_{Gd-Gd}$ and $d_{Co-Co}$) that are positive, two scenarios of AFM DMI ($d_{Gd-Co}$) are considered. At 300 K, in all thicknesses, larger DMI is needed to form skyrmions with positive $d_{Gd-Co}$ than with negative $d_{Gd-Co}$. In 5 nm sample, D = 0.55 mJ/m$^2$ is needed to stabilize skyrmions with $d_{Gd-Co} > 0$. In comparison, with $d_{Gd-Co} < 0$, a smaller DMI of D = 0.21 mJ/m$^2$ is needed to stabilize skyrmions. Similar behaviors are also found in 10 nm and 15 nm samples. With $d_{Gd-Co} > 0$, the smallest skyrmions are found at D = 1.26 mJ/m$^2$ in 10 nm sample and D = 2.31 mJ/m$^2$ in 15 nm sample. On the other hand, with $d_{Gd-Co} < 0$, the smallest skyrmions are found at D = 0.84 mJ/m$^2$ in 10 nm sample and D = 1.68 mJ/m$^2$ in 15 nm sample.

To understand such intriguing behavior in a FiM, the in-plane spin configurations and the chirality of the skyrmion wall are investigated. **Fig. 5** summarizes the chirality of the skyrmion walls in the Co sublattice. Using $d_{Gd-Gd}$, $d_{Co-Co} > 0$ and $d_{Gd-Co} < 0$, in the Co sublattice, the spins are turning in counter-clockwise direction across the skyrmion wall. For Gd sublattice, the spins in the skyrmion wall are also turning counter-clockwise. This can be explained by the DMI in the system. AFM couplings between Gd and Co align the spins of Gd and Co in nearly antiparallel directions, except a small canting due to the presence of DMI. With $d_{Gd-Gd}$ and $d_{Co-Co} > 0$, turning counter-clockwise is energetically favorable. However, with $d_{Gd\text{-}Gd}$, $d_{Co\text{-}Co} > 0$ and $d_{Gd\text{-}Co} > 0$, the chirality of the simulated skyrmion wall is found to be opposite. The DMI torque between the AFM pairs now opposes the DMI torques within each sublattice. In the presence of a stronger inter-sublattice DMI torque, the spins in each sublattice now turn clockwise across the skyrmion wall.

To better illustrate the change in chirality, the total DMI energies between Co-Co, Gd-Gd and Gd-Co are computed using the equilibrium configurations at 0 K. **Table 1** summarizes the sign of the total DMI energies for different nearest neighbor pairs. With $d_{Gd\text{-}Gd}$, $d_{Co\text{-}Co} > 0$ and $d_{Gd\text{-}Co} > 0$, spins are turning counter-clockwise. With this configuration, the total DMI energy between Gd-Gd pair $E_{DMI}$(Gd-Gd) and Co-Co pair $E_{DMI}$(Co-Co) are negative, and the total DMI energy between Gd and Co pair $E_{DMI}$(Gd-Co) is also negative. This means that with $d_{Gd\text{-}Gd}$, $d_{Co\text{-}Co} > 0$, it is energetically favorable for spins to turn counterclockwise. On the other hand, with $d_{Gd\text{-}Gd}$, $d_{Co\text{-}Co} > 0$ and $d_{Gd\text{-}Co} > 0$, spins are revealed to turn clockwise from the simulated configurations. As a result of the sign change in chirality, $E_{DMI}$(Gd-Gd) and $E_{DMI}$(Co-Co) become positive. On the other hand, $E_{DMI}$(Gd-Co) remains negative, because both chirality and $d_{Gd\text{-}Co}$ changes sign. This implies that it is energetically favorable for Gd-Co pair to turn clockwise across, but it is energetically unfavorable for Gd-Gd and Co-Co pairs to do so. In other word, AFM DMI $d_{Gd\text{-}Co}$ is able to overcome ferromagnetic DMI $d_{Gd\text{-}Gd}$ and $d_{Co\text{-}Co,}$ resulting in energy favorable configurations for Gd-Co pairs. To summarize, in a FiM, if the DMI of ferromagnetic pair and AFM pair have the same sign, a cancellation of DMI occurs because it is preferable for a ferromagnetic pair to turn in the opposite direction of an AFM pair. No cancellation occurs if the DMI of ferromagnetic pair and AFM pair have the opposite sign. These also explain the differences in size of skyrmion between $d_{Gd\text{-}Co} < 0$ and $d_{Gd\text{-}Co} > 0$. The $d_{Gd\text{-}Co} < 0$ scenario has larger skyrmions because both ferromagnetic and AFM pairs are contributing to the formation of a skyrmion, which means the DMI effect is stronger overall.

To investigate the minimal size of room temperature skyrmions in GdCo, D-K phase diagrams with exponentially decaying DMI at 300 K are simulated for 5, 10 and 15 nm GdCo films. In this section, we focus on the $d_{Gd\text{-}Co} < 0$ scenario. Since energy barrier is a function of exchange stiffness and thickness[18], the minimal skyrmions size found in $d_{Gd\text{-}Co} < 0$ scenario can also apply to $d_{Gd\text{-}Co} > 0$ scenario, except a larger DMI is required. For each thickness, anisotropy ranges from 0.05 x $10^5$ J/m$^3$ to 4 x $10^5$ J/m$^3$ are investigated. Experimentally, GdCo has anisotropy in the order of $10^4$ J/m$^3$.[25,36] For DMI, larger interfacial DMI is explored in thicker samples, because as thickness increases, the average DMI decreases, and larger interfacial DMI is needed to stabilize skyrmions. In 5 nm samples, interfacial DMI of 0 to 2 mJ/m$^2$, which corresponds to $d_{Co\text{-}Co}$ of 0 to 2.38 x $10^{22}$ J, are investigated. **Fig. 6 (a)** shows the D-K phase diagram of 5 nm GdCo at 300 K. In 5 nm GdCo, skyrmions range from 12 nm to 40 nm are stabilized in the simulated range of interfacial DMI and anisotropy. Lines of 15 to 30 nm indicate the size of skyrmions at various DMI and anisotropy. As DMI decreases or anisotropy increases, skyrmions become smaller and eventually collapse into FiM states. At the opposite side of D-K diagram, with large DMI and small anisotropy, skyrmions larger than 40 nm becomes elongated or collapsed due to the boundary of the simulation space (50.7 nm x 50.7 nm). This elongation of skyrmions was also seen earlier in **Fig. 4** at large DMI values. Overall, for a given anisotropy, as interfacial DMI increases from 0 to 2.0 mJ/m$^2$, the equilibrium configuration goes from FiM to skyrmions, then to stripes. For a fixed DMI, as anisotropy increases, size of skyrmions decreases, and finally, skyrmions collapse into FiM states. These behavior of skyrmions in FiM GdCo as a function of DMI and anisotropy is the same as what has been observed in a ferromagnet[17,18].

For 10 nm and 15 nm GdCo, DMI of 0 to 3 mJ/m$^2$ ($d_{Co-Co}$ of 0 to 3.57 x $10^{22}$ J) and 0 to 4 mJ/m$^2$ ($d_{Co-Co}$ of 0 to 4.76 x $10^{22}$ J) are explored respectively. The overall trend of skyrmions as a function of DMI and anisotropy in 10 nm and 15 nm GdCo are identical to that of 5 nm GdCo, where increase in DMI leads to larger skyrmions, and increase in anisotropy results in smaller skyrmions. However, one difference in thicker samples from 5 nm sample is that ultrasmall skyrmions as small as 7 nm are stable in room temperature. For both 10 nm and 15 nm GdCo, there is a region of DMI and anisotropy where ultrasmall skyrmions are stablized. In 10 nm GdCo, ultrasmall skyrmions are found in the region of DMI ranges from 0.8 to 1.0 mJ/m$^2$ and anisotropy ranges from (0.1 to 0.8) x $10^5$ J/m$^3$. For 15 nm GdCo, this region lays within DMI ranges from 1.5 to 1.8 mJ/m$^2$ and anisotropy ranges from (0.1 to 1.0) x $10^5$ J/m$^3$. For both 10 nm and 15 nm GdCo, the anisotropy falls within the same range as what has being measured experimentally in GdCo[25,36], which is in the order of $10^4$ J/m$^3$. However, the interfacial DMI is less than what has been observed at a Pt interface. Ab-inito calculation has found Interfacial DMI of up to 12 mJ/m$^2$ is reported at an ideal Pt/Co interface[30]. On the other hand, the interfacial DMI measured in Co/Pt and other Co-alloy/Pt films are around 1.2 to 1.5 mJ/m$^2$. [52-54] Thus, some reductions of DMI from that of Pt are needed to experimentally obtain ultrasmall skyrmion in 5 and 10 nm GdCo films. Reduction of DMI can be obtained by sandwiching GdCo between two Pt layers with one Pt layer being diluted by other elements. Since GdCo is amorphous, we have more flexibility of tuning the underlayer and the capping layer of a multilayer sandwich. With its intrinsic anisotropy and flexibility, GdCo films are promising materials to obtain ultrasmall skyrmions at room temperature through DMI tuning.

For device applications, especially in thicker films, we will also need to consider the growth of skyrmions away from the interface. With decaying DMI away from the interface, spins at the top of a thicker sample experience effectively zero DMI. Without DMI, one might expect spins near the top to align parallel for FM neighbors and antiparallel for AFM neighbors, and skyrmions to disappear far away from the interface. If skyrmions collapse far away from the interface, the reliability of such memory devices would be vastly reduced. To investigate whether skyrmions remain robust in thicker samples, a numerical tomography is employed to image simulated ultrasmall skyrmions at 300 K. **Fig. 7** shows the numerical tomography plot of a ultrasmall skyrmion in 10 nm GdCo. This skyrmion corresponds to D = 0.84 mJ/m$^2$ and K = 0.3 x $10^5$ J/m$^3$. The same skyrmion was shown in **Fig. 4(b)** and as the smallest skyrmions (Star Symbol) at K = 0.3 x $10^5$ J/m$^3$ in **Fig. 6(b)**. In the 3D plots at the center of **Fig. 7**, colors are made to be somewhat transparent to reveal the skyrmions structure near the center. For Co sublattice, red to orange color shows that most of the spins are pointing down. A region of green and blue that appears near the center corresponds to the simulated skyrmion at 300 K. As evidenced by the columnar distribution of blue color, the skyrmion retains a uniform columnar growth from the bottom to the top. Columnar distribution of skyrmion is also found in Gd sublattice, where a column of red is distributed uniformly from the bottom to the top. This feature can be understood in terms of the large magnetic exchange length >20 nm due to the low magnetization in the ferrimagnet.

To further demonstrate the uniform columnar distribution of skyrmion, in-plane and out-of-plane cross sections of the skyrmion are also plotted in **Fig. 7**. On the left of **Fig. 7**, in-plane cross section of spin configuration within 0.5 nm of the interface and 0.5 nm of the top are mapped. The skyrmions at the interface and near the top have identical size and shape. Compare to the mapping of spin configurations in **Fig. 4(b)**, size of the skyrmion remain the same. This shows that the size of skyrmions remain the uniform throughout a sample. On the right side of **Fig. 7**, out-of-plane cross sections are shown for Gd and Co sublattices. The blue color in Co sublattice and the red color in Gd sublattice correspond to the center of the skyrmion. For both sublattice, out-of-plane cross sections show a columnar distribution of skyrmion from the bottom interface to the top. These results provide important evidences that skyrmion remain robust through a thicker sample, and further support of using thicker GdCo samples to increase skyrmion stability at room temperature.

**Conclusions**

Using atomistic stochastic LLG simulations, ultrasmall skyrmions are shown to be stable at room temperature in ferrimagnetic GdCo. Despite the rapid decay of Dzyaloshinskii Moriya interaction (DMI) away from the interface, a realistic range of DMI values is seen to stabilize skyrmions in GdCo films as thick as 15 nm irrespective of the sign of DMI between antiferromagnetic coupled Gd and Co, Furthermore, the low DMI values needed to form ultrasmall skyrmion in GdCo indicate opportunity for designing magnetic materials to host ultrasmall Neel skyrmions. Through tomography of an ultrasmall skyrmion in 10-nm thick GdCo film, it is discovered that the skyrmion assumes a columnar configuration that extends uniformly across the film thickness despite having near zero DMI far away from the interface. These findings argue for using thicker magnetic films to host ultrasmall skyrmions, providing an important strategy for developing high density and high efficiency skyrmion based devices.

**Methods**

The classical atomistic Hamiltonian H in Eq. (1) is employed to investigate magnetic textures in amorphous FiMs.

$$H = -\frac{1}{2}\sum_{<i,j>} J_{ij}\boldsymbol{s_i}\cdot\boldsymbol{s_j} - \frac{1}{2}\sum_{<i,j>} D_{ij}\cdot\left(\boldsymbol{s_i}\times\boldsymbol{s_j}\right) - K_i\left(\boldsymbol{s_i}\cdot\widehat{\boldsymbol{K_i}}\right)^2$$

$$-\mu_0\mu_i\boldsymbol{H_{ext}}\cdot\boldsymbol{s_i} - \mu_0\mu_i\boldsymbol{H_{demag}}\cdot\boldsymbol{s_i} \quad (2)$$

where $\boldsymbol{s_i}, \boldsymbol{s_j}$ are the normalized spins and $\mu_i, \mu_j$ are the atomic moments at sites i, j respectively. The atomic moment is absorbed into the exchange constant, $J_{ij} = \mu_i\mu_j j_{ij}$, the DMI interaction $\boldsymbol{D_{ij}} = \mu_i\mu_j\boldsymbol{d_{ij}}$, which is proportional to $\mathbf{r}_i \times \mathbf{r}_j$, the positional vector between the atoms i, j and the interface, and the effective anisotropy $K_i = \mu_i k_i$. $\boldsymbol{H_{ext}}$ and $\boldsymbol{H_{demag}}$ are the external field and demagnetization field respectively.

Only nearest neighbor interactions are considered in the exchange and DMI interactions. Periodic boundary conditions are enforced in the x and y directions.

To find the ground state, the spins are evolved under the following stochastic Landau-Lifshitz-Gilbert (LLG) equation,

$$\frac{d\boldsymbol{M}}{dt} = -\frac{\gamma}{1+\alpha^2}\boldsymbol{M}\times\left(\boldsymbol{H_{eff}}+\boldsymbol{\xi}\right) - \frac{\gamma\alpha}{(1+\alpha^2)M_s}\boldsymbol{M}\times\left[\boldsymbol{M}\times\left(\boldsymbol{H_{eff}}+\boldsymbol{\xi}\right)\right] \quad (3)$$

where $\gamma$ is the gyromagnetic ratio, $\alpha$ is the Gilbert damping constant, $\boldsymbol{H_{eff}}$ is the effective field, $\boldsymbol{\xi}$ is the Gaussian white noise term for thermal fluctuations and $M_s$ is the saturation magnetization.

The parameters used in our simulation are listed in **Table 2**. Exchange couplings $J_{ij}$ are calibrated based on Ostler *et al.*[49] to maintain the same Curie temperature and compensation temperature for a given compensation. At 300 K, the magnetization of $Gd_{25}Co_{75}$ is 5 x $10^4$ A/m. Anisotropy energy is determined based on Hansen *et al.*[36]

To incorporate the amorphous short range order, an amorphous structure of a 1.6 nm x 1.6 nm x 1.6 nm box containing 250 atoms is generated from *ab initio* molecular dynamics calculations by Sheng *et al*.[58]. The composition used in the simulation is $Gd_{25}Co_{75}$. **Fig. 1** shows a plot of RE and TM atoms in the amorphous structure. For a 4.8 nm thick sample, replicas of this box (32 x 32 x 3) are placed next to each other to expand the simulated sample to 50.7 nm x 50.7 nm x 4.8 nm and 768000 atoms. On average, we find that each Co atom has 6.8 Co neighbors and 4.1 Gd neighbors, while each Gd atom has 11.7 Co neighbors and 3.5 Gd neighbors. We have also employed a FCC lattice to study skyrmions in GdCo. We found that with the same compensation temperature and magnetization, a larger DMI is needed to stabilize skyrmion in a FCC lattice structures than amorphous structure. This is because the overall effectiveness of DMI is affected by the structure. Only results using the amorphous structure are shown herein.

In the simulations, the initial states are skyrmion of 20 nm based on the 2-pi model[18]. Various initial states, includes random initial states and 10-30 nm skyrmions, have been tested and found to produce the same final states. The size of skyrmions are defined as the diameter for which $M_z$ = 0. Since skyrmions are not perfectly symmetric, size of skyrmion is the average diameter.

**Data Availability**

The datasets generated during and/or analyzed during the current study are available from the corresponding author on reasonable request.

**Acknowledgements:**

This work was supported by the DARPA Topological Excitations in Electronics (TEE) program (grant D18AP00009). The content of the information does not necessarily reflect the position or the policy of the Government, and no official endorsement should be inferred. Approved for public release; distribution is unlimited. This work was partially supported by NSF-SHF-1514219.

**Additional Information**

**Author Contributions**

S.J.P. conceived the project and supervised the simulation, C.T.M. performed the simulation, Y.X. assisted in the simulation and provided comments, A.W.G. discussed skyrmion stability and suggested improvement to the manuscript, H.S. provided the amorphous structure from ab initio molecular dynamics calculations.

**Competing Interests**

The authors declare no competing interests.

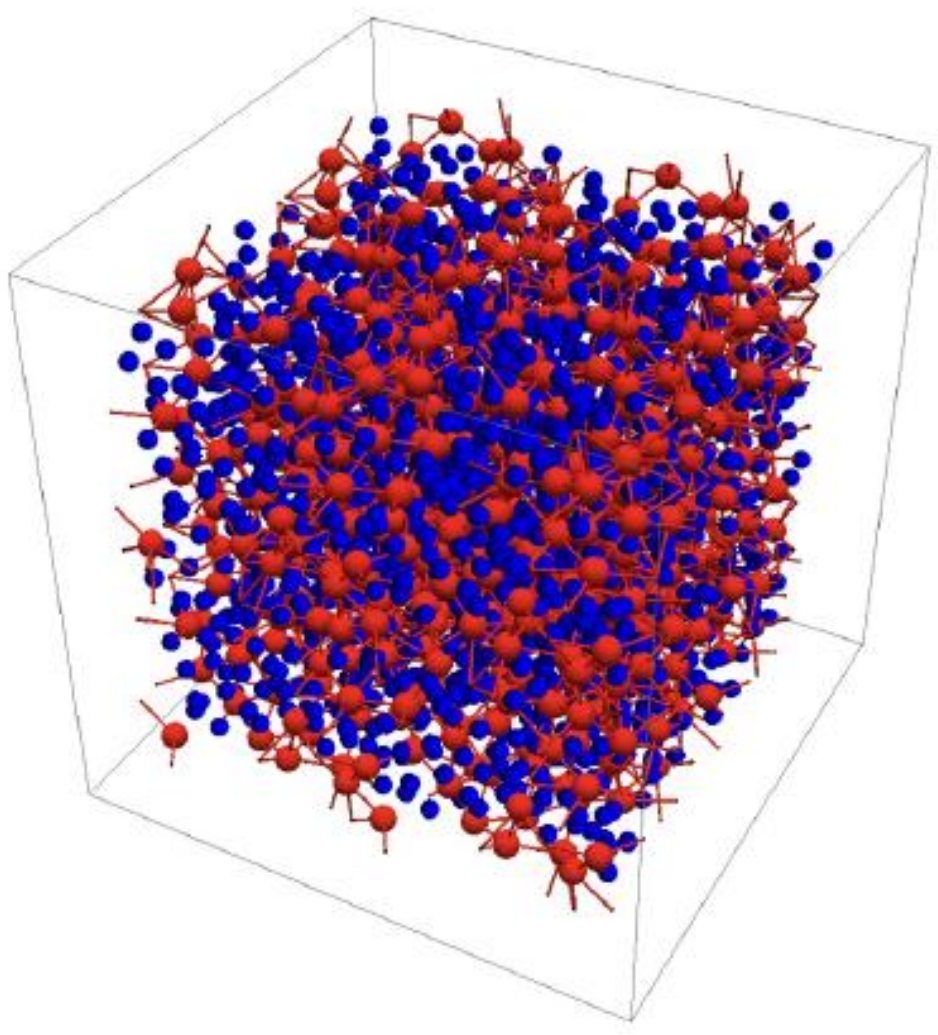

**Figure 1 Amorphous structure of $RE_{25}TM_{75}$ from ab initio molecular dynamics calculations**. Red atoms are rare-earth, and blue atoms are transition-metal.

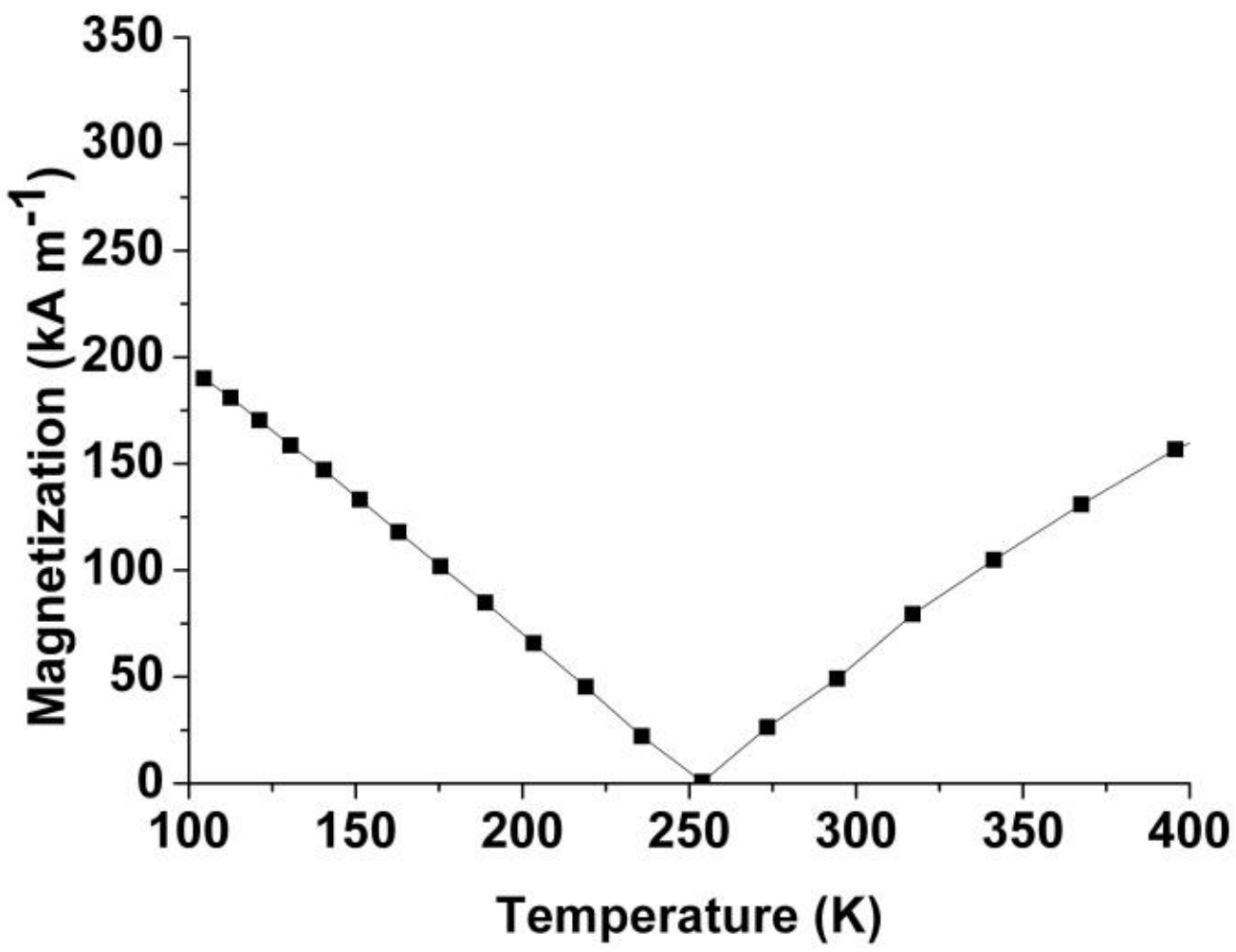

**Figure 2 Simulated saturation magnetization vs. temperature of amorphous $Gd_{25}Co_{75}$.** The compensation temperature of amorphous $Gd_{25}Co_{75}$ is near 250 K, and the magnetization is small at room temperature.

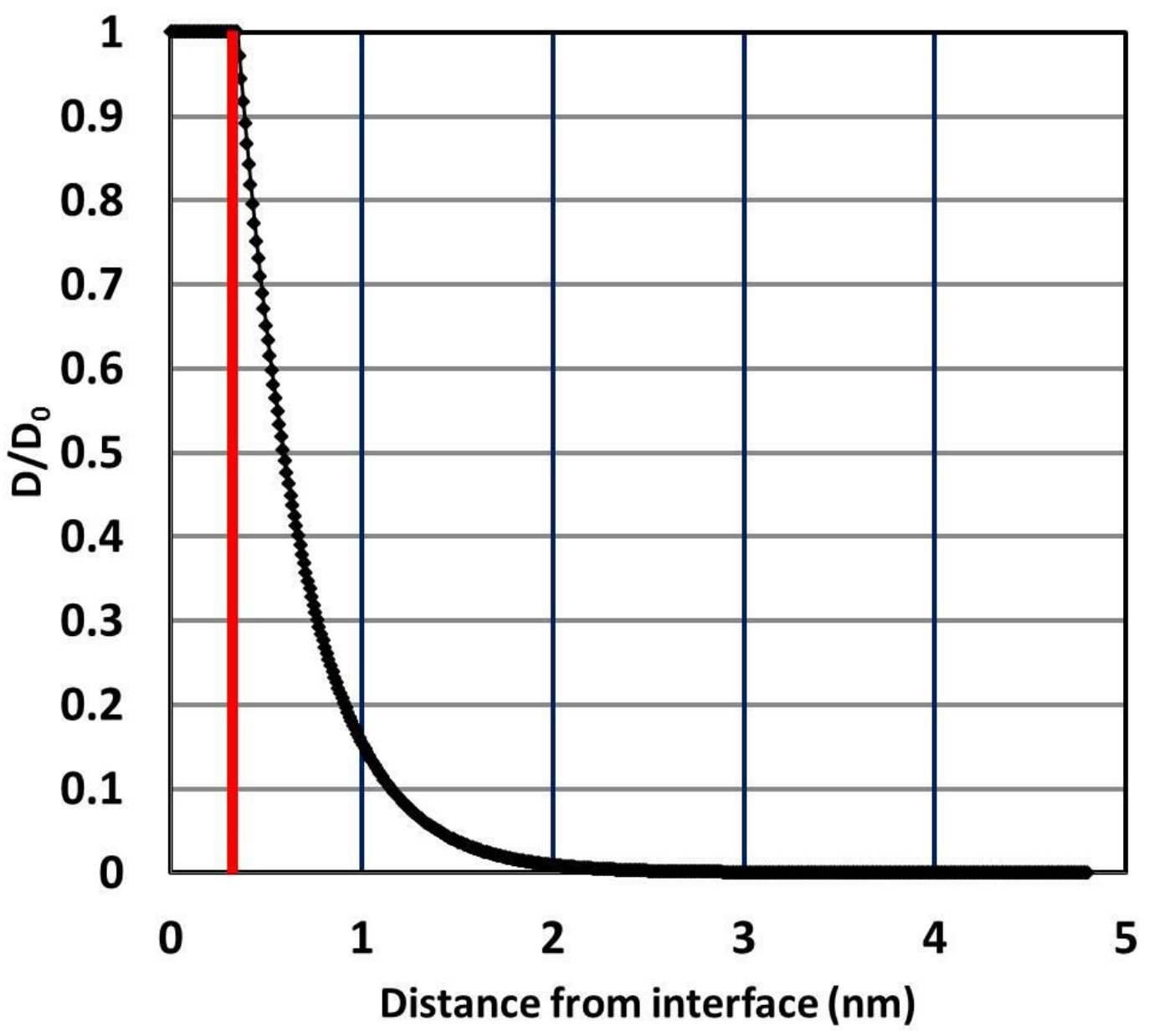

**Figure 3 Exponential decay DMI in 5 nm GdCo as function of distance from bottom interface (z).** In this model, DMI remains constant ($D_0$) within 0.35 nm of the bottom interface, as indicated by the red line. Away from the interface, the strength of DMI decays exponentially as shown.

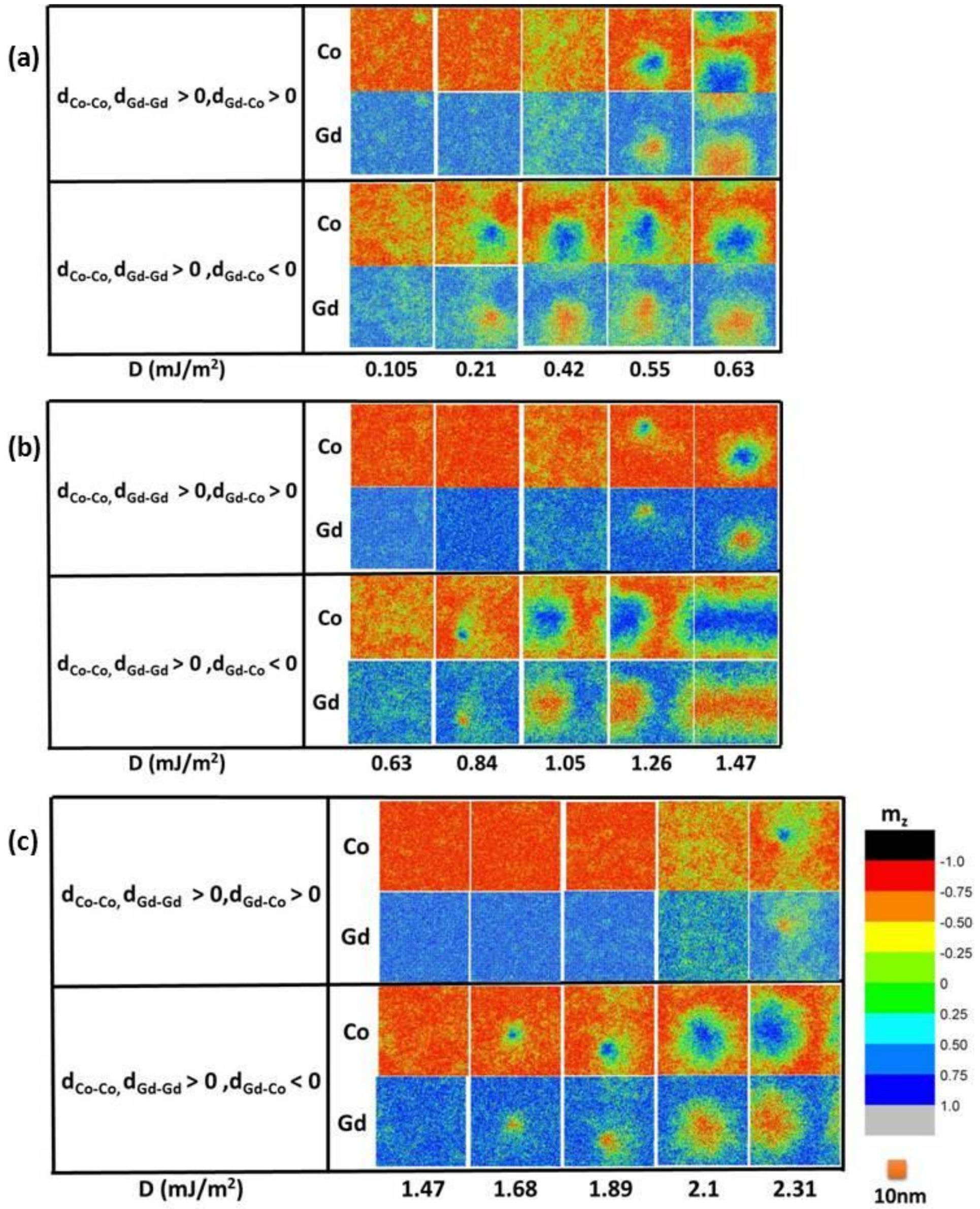

**Figure 4 Color mapping of equilibrium spin configurations for various DMI values (exponentially decaying DMI) at 300K for $d_{Gd\text{-}Co} > 0$ and $d_{Gd\text{-}Co} < 0$ in (a) 5 nm, (b) 10 nm, and (c) 15 nm GdCo.** Out-of-plane components of reduced magnetizations ($m_z$) are mapped in the x-y plane using the color bar shown in (c). Ultrasmall skyrmions are revealed in 10 nm and 15 nm GdCo samples.

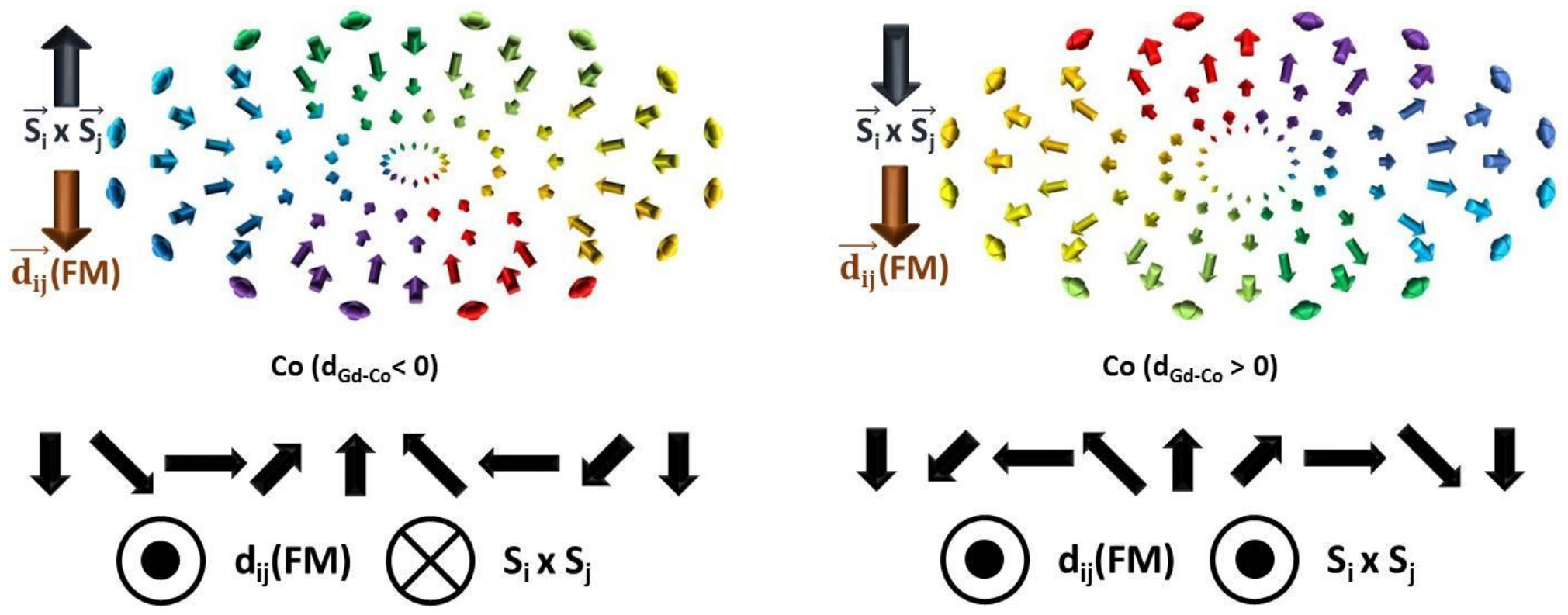

**Figure 5 Simulated skyrmion configurations of Co sublattice for $d_{Gd-Co} < 0$ and $d_{Gd-Co} > 0$ with metal interface at the bottom. (Top) An overhead view of simulated skyrmion configurations for $d_{Gd-Co} < 0$ and $d_{Gd-Co} > 0$. (Bottom)** For $d_{Gd-Co} < 0$, the skyrmion wall is turning counter-clockwise. The **d**(FM) vector is pointing in the opposite direction of $\mathbf{S_i} \times \mathbf{S_j}$. $E_{DMI}$ (FM)= $\mathbf{d_{ij}} \cdot (\mathbf{S_i} \times \mathbf{S_j})$ is negative, which is favorable. For $d_{Gd-Co} > 0$, the skyrmion wall is turning clockwise. The **d**(FM) vector and $\mathbf{S_i}$ x $\mathbf{S_j}$ are pointing in the same direction, resulting in positive $E_{DMI}$ (FM). Identical signs of the DMI energy are also found in the Gd sublattice.

| Scenario | $E_{DMI}$(Gd-Gd) | $E_{DMI}$(Co-Co) | $E_{DMI}$(Gd-Co) |
|---|---|---|---|
| $d_{Gd-Gd}$, $d_{Co-Co}$ **> 0,** $d_{Gd-Co} < 0$ | - | - | - |
| $d_{Gd-Gd}$, $d_{Co-Co}$ **> 0,** $d_{Gd-Co} > 0$ | + | + | - |

**Table. 1 Sign of total DMI energy $E_{DMI}$ computed from equilibrium spin configurations at 0 K.**

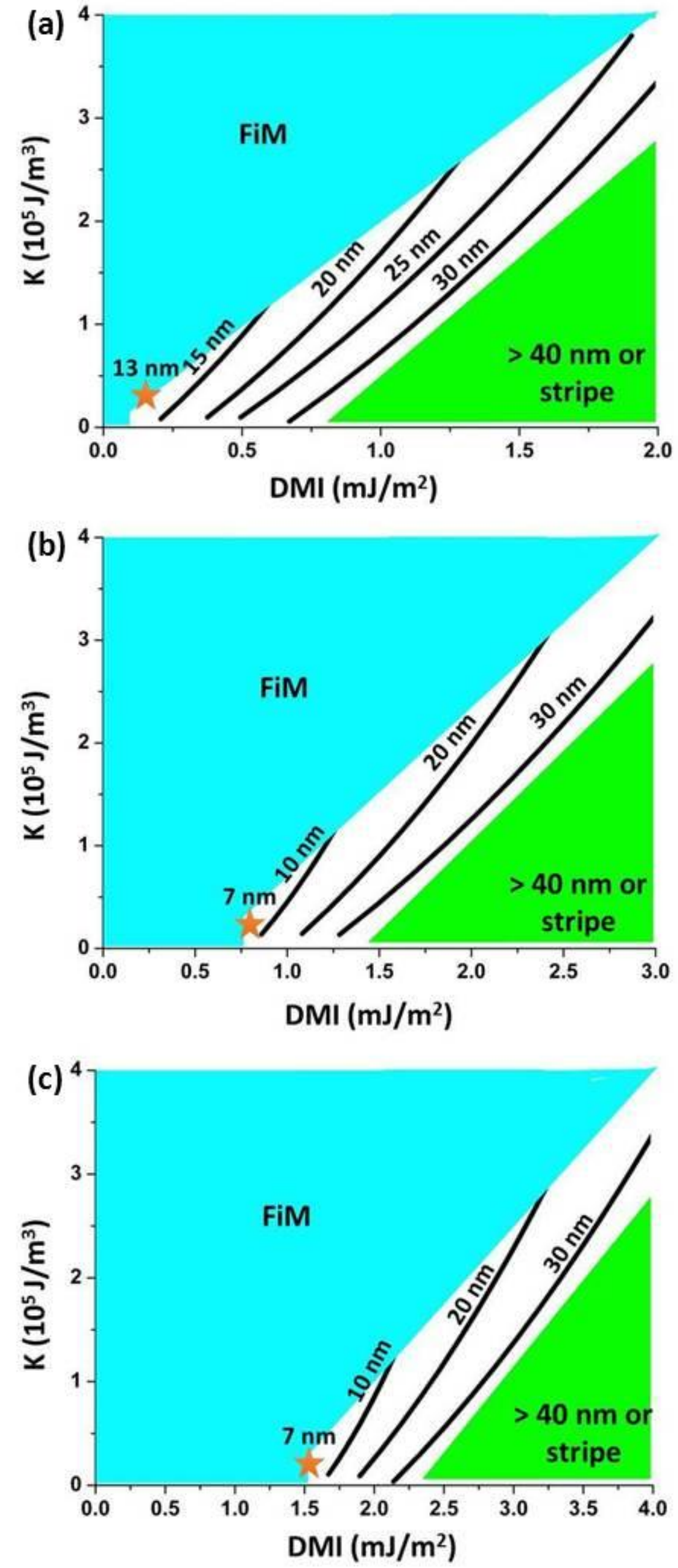

**Figure 6 D-K phase diagram of (a) 5 nm, (b) 10 nm and (c) 15 nm GdCo at 300 K with $d_{Gd\text{-}Co} < 0$. Star corresponds to smallest skyrmions simulated at K = 0.3 x $10^5$ J/m$^3$.** Ultrasmall skyrmions are revealed in 10 nm and 15 nm GdCo. Due to limits of simulation space (50.7 nm x 50.7 nm), with periodic boundary conditions in x-y direction, large skyrmions (> 40 nm) become either elongated or collapsed.

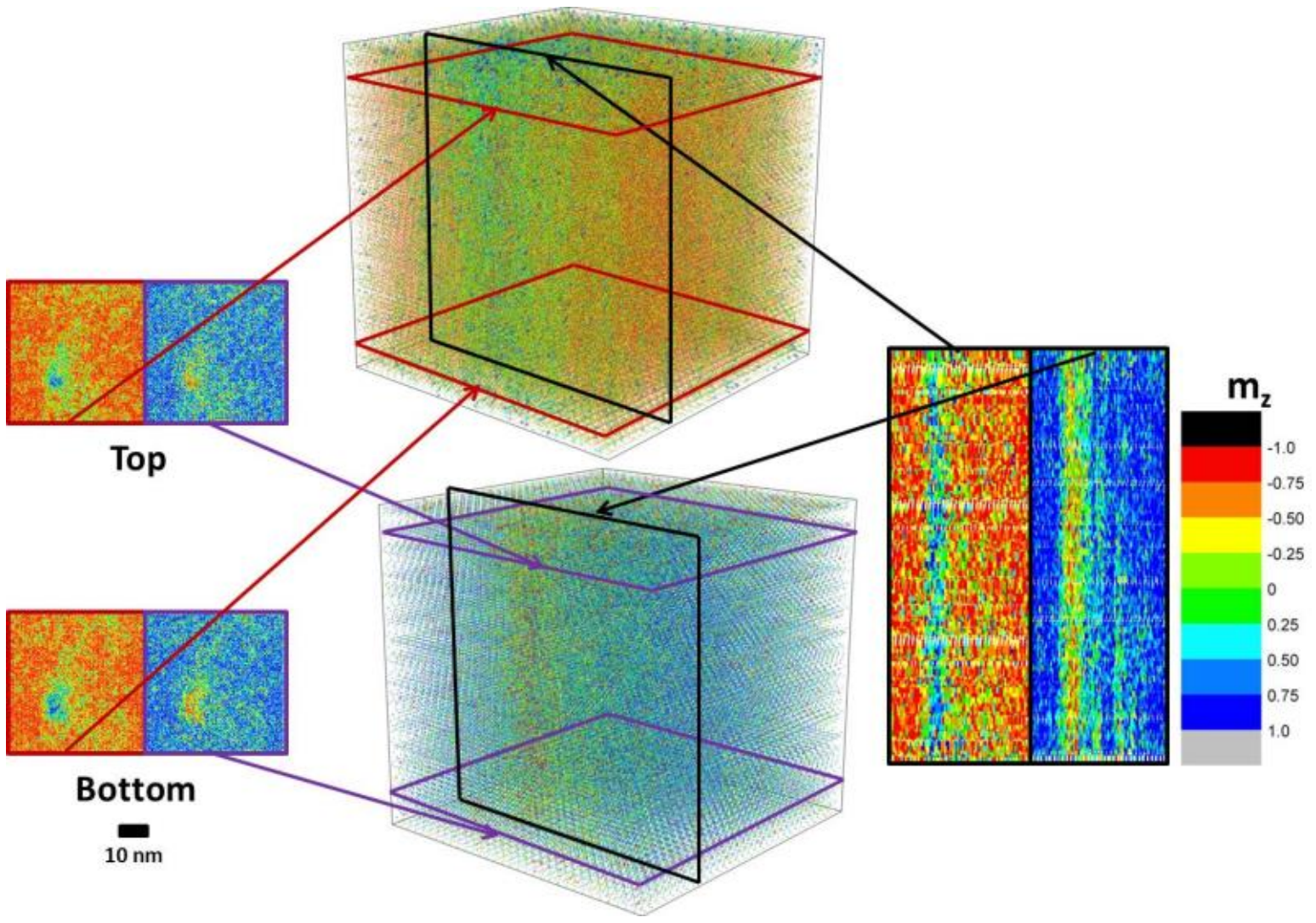

**Figure 7 Tomograph of a simulated ultrasmall skyrmion in 10 nm GdCo at 300 K.** It reveals columnar skyrmion distribution throughout the 10 nm GdCo sample. The figure shows Co-sublattice spins (top box), Gd-sublattice spins (bottom box), in-plane cross sections of near the top and bottom interface (left), and out-of-plane cross sections (right).

| Parameter | **Value** |
|---|---|
| **Gyromagnetic ratio (ϒ)** | 2.0023193 |
| **Gilbert Damping (α)** | 0.05 |
| **Gd moment ($\mu_{Gd}$)** | 7.63 $\mu_B$ |
| **Co moment ($\mu_{Co}$)** | 1.72 $\mu_B$ |
| **Gd-Gd exchange constant ($J_{Gd-Gd}$)** | 1.26 x $10^{-21}$ J |
| **Co-Co exchange constant ($J_{Co-Co}$)** | 3.82 x $10^{-21}$ J |
| **Gd-Co exchange constant ($J_{Gd-Co}$)** | -1.09 x $10^{-21}$ J |
| **Magnetic Field (H)** | 0.01 T |

**Table. 2 Values of parameters used in the simulation.**